\documentstyle[12pt]{article}

\advance\hoffset by -0.5cm
\def\3{\ss}
\def\sq{\hbox{\rlap{$\sqcap$}$\sqcup$}}
\def\qed{\ifmmode\sq\else{\unskip\nobreak\hfil
\penalty50\hskip1em\null\nobreak\hfil\sq
\parfillskip=0pt\finalhyphendemerits=0\endgraf}\fi}
\def\half {\frac{1}{2}}
\def\thalf {\frac{3}{2}}
\def\ct{\tilde{c}}
\def\bbbz {{\sf Z\!\!Z}}

\def\bbbn {{\rm I\!N}}
\def\bbbone {{\mathchoice {\rm 1\mskip-4mu l} {\rm 1\mskip-4mu l}
{\rm 1\mskip-4.5mu l} {\rm 1\mskip-5mu l}}}
\def\bbbc{{\mathchoice {\setbox0=\hbox{$\displaystyle\rm C$}\hbox{\hbox
to0pt{\kern0.4\wd0\vrule height0.9\ht0\hss}\box0}}
{\setbox0=\hbox{$\textstyle\rm C$}\hbox{\hbox
to0pt{\kern0.4\wd0\vrule height0.9\ht0\hss}\box0}}
{\setbox0=\hbox{$\scriptstyle\rm C$}\hbox{\hbox
to0pt{\kern0.4\wd0\vrule height0.9\ht0\hss}\box0}}
{\setbox0=\hbox{$\scriptscriptstyle\rm C$}\hbox{\hbox
to0pt{\kern0.4\wd0\vrule height0.9\ht0\hss}\box0}}}}

\begin{document}
\thispagestyle{empty}
\begin{flushright}
DAMTP-93-56 \\
hep-th/9309105
\end{flushright}
\begin{center}
\vspace{2.5cm}

{\huge Fusion rules of chiral algebras}
\vspace{1.0cm}

{\large Matthias Gaberdiel} \footnote{e-mail: M.R.Gaberdiel@amtp.cam.ac.uk} \\
{Department of Applied Mathematics and Theoretical
Physics\\
University of Cambridge, Silver Street \\
Cambridge, CB3 9EW, U.\ K.\ }
\vspace{0.5cm}

{September 1993}
\vspace{0.5cm}

{Abstract}
\end{center}

Recently we showed that for the case of the WZW- and
the minimal models fusion can be understood as a certain
ring-like tensor product of the symmetry algebra.
In this paper we generalize this analysis to arbitrary
chiral algebras. We define the tensor product
of conformal field theory in the general case and prove that
it is associative and symmetric
up to equivalence. We also determine explicitly
the action of the chiral algebra on this tensor product.
\smallskip

In the second part of the paper we demonstrate that
this framework provides a powerful tool for calculating
restrictions for the fusion rules of chiral algebras. We exhibit this
for the case of the $W_{3}$-algebra
and the $N=1$ and $N=2$ NS superconformal algebras.

\section{Introduction}

Fusion is a central concept in conformal field theory.
The fusion rules of a conformal field theory describe
which of the three-point-functions of the theory
are non-zero and thus determine the possible couplings.
Equivalently, the fusion rules can be understood
to describe which conformal families appear in the operator
product expansion of two vertex operators.

This latter point of view suggests that
one should regard fusion as
some kind of tensor product. Conformal families
can be interpreted as irreducible representations
of the chiral algebra and multiplying vertex operators
should correspond to taking the tensor product.
The operator product expansion is then the decomposition
of this tensor product into the irreducible components.

Generically, the chiral algebra (of which this
tensor product has to be a representation) possesses
a central extension and thus there is no canonical
definition of a tensor product. However, inspired
by a recent proposal of Richard Borcherds \cite{B}
to regard fusion in conformal field theory as the
canonical tensor product of modules of a {\em quantum
ring}, a generalization of rings and vertex
algebras,  we showed in \cite{MG} how one can
define a tensor product for the case of the WZW- and
the minimal models. We determined explicitly
the action of the chiral algebra on this tensor product
and showed that the tensor product is
associative and symmetric up to equivalence.
We then determined the fusion rules of these models,
analyzing the decomposition of the tensor product
into its irreducible components. We thereby recovered
the well-known restrictions for the fusion rules.
\medskip

In this paper we want to generalize this analysis
to arbitrary chiral algebras. In section 2 we
derive the action of a general chiral algebra on products
of vertex operators, which gives rise to an action of
the chiral algebra on tensor products
of representations. This action can be formulated as
a comultiplication which depends on two parameters, namely the
insertion points of the two vertex operators. For certain values of
these parameters we have two different expressions, which
agree on the underlying conformal field theory.
In order to maintain the whole information of conformal
field theory, we have to {\em impose} this equality
on the level of representations. This leads us
to defining the ``true'' tensor product of conformal field
theory as the quotient of the vector space tensor product
by all relations which arise in this way.

The action of the chiral algebra on this tensor product is
given by either of the two expressions.
As in \cite{MG} we can prove, that the so-defined tensor product
is associative and symmetric up to equivalence.
There we were also able to show, that these expressions
do indeed define a comultiplication. We cannot
repeat this proof here, as we have not specified the
commutation (or anticommutation) relations
of the chiral algebra.
However, we can show on general grounds that they have to
satisfy this property if they come from a well-defined
conformal field theory.
\smallskip

We have thus established a framework within which fusion
is a well-defined algebraic notion. Determining the fusion rules
now amounts to decomposing this tensor product
into irreducible representations. We cannot say very much
about this in the general case, however, given any
algebra we can use the knowledge about the null-vectors
to determine restrictions for the possible fusion rules.
In particular, as we have {\em proved} the associativity of
the tensor product, we only have to
analyze the fusion of simple representations to
obtain restrictions for the general case. We use this
argument to derive restrictions for the fusion rules of the
$W_{3}$-algebra and the $N=1$ and $N=2$ NS superconformal
algebras in section 3-5, thereby reproducing the
results of \cite{FZ}, \cite{SS} and \cite{MSS1}.
Our method, however, is entirely different: it is purely
algebraic and does not (for the case of
the superconformal algebras) rely on any superdifferential
calculus.

\section{Definition of the tensor product}
\renewcommand{\theequation}{2.\arabic{equation}}
\setcounter{equation}{0}
\setcounter{footnote}{0}

Let us start by defining the chiral algebra of a
conformal field theory. Let $S(w)$ be a holomorphic
field with conformal weight $h\in\bbbz /2$, where
$L_{0}\, S(0)\, \Omega = h\, S(0)\, \Omega$.
Following \cite{Peter89} we can expand $S$
in terms of modes as
\begin{equation}
\label{mode}
S(w)=\sum_{l\in\bbbz + h} w^{l-h} \; S_{-l}.
\end{equation}
The operator product expansion of two holomorphic fields
contains only holomorphic fields. We thus define
the chiral algebra ${\cal A}$ to be
the algebra generated by all the modes of all holomorphic fields,
subject to the commutation (or anticommutation) relations
induced from the (singular part of the) operator product expansion.
This definition contains most ``symmetry algebras'' of
conformal field theories, but not for example
the Ramond superconformal algebras. On the other hand
we do not want to include the Ramond algebras in our
definition, as the operator product expansion of two Ramond fields
contains only Neveu-Schwarz fields and this does not allow
an interpretation as a tensor product.\\
The operator product expansion of a holomorphic
field $S$ with a vertex operator
is then given by \cite{Peter89}
\begin{equation}
\label{ope}
S(w) V(\psi,\zeta)=\sum_{l\in\bbbz +h} (w-\zeta)^{l-h}
V(S_{-l}\psi,\zeta).
\end{equation}
\smallskip

We want to determine the action of the modes $S_{l}$ on
products of vertex operators. To do this we have
to calculate (as in \cite{MG}) the contour integral
\begin{equation}
\label{int}
\oint_{C} dw \; w^{l+h-1} \;S(w) \;V(\psi,\zeta) V(\chi,z)\;\Omega ,
\end{equation}
where $C$ is a large contour encircling the two insertion
points and we assume $|\zeta|>|z|$ in order to guarantee
the existence of the operator product.
We could use the usual contour integration techniques to
evaluate (\ref{int}), as the residues of the integrand
are determined by (\ref{ope}). However, if we did this
calculation straight away, we would obtain terms, where
the $S$-modes act on the vacuum --- and these terms would
not allow an interpretation as a comultiplication.
Instead we consider scalar products of the integrand
of (\ref{int}) with any vector in the dense
subset of finite energy vectors $\varphi\in{\cal F}$ and
obtain thus (by the definition of conformal field
theory) a meromorphic function of $w$
\begin{equation}
\label{mer}
\left\langle\varphi, S(w) \;V(\psi,\zeta)
V(\chi,z)\;\Omega\right\rangle,
\end{equation}
whose singular part is given by
\begin{equation}
\label{sing}
\varepsilon_{\psi}
\sum_{l=-\infty}^{h-1} (w-z)^{l-h}
\langle\varphi, V(\psi,\zeta) V(S_{-l}\chi,z)\;\Omega\rangle +
\sum_{m=-\infty}^{h-1} (w-\zeta)^{m-h} \langle\varphi,V(S_{-m}\psi,\zeta)\;
V(\chi,z)\;\Omega\rangle.
\end{equation}
Here $l$ and $m$ are in $\bbbz +h$ and we assume
that the vertex operator $V(\psi,\zeta)$ and
the holomorphic field $S^{a}(w)$ are local with respect to
each other, i.\ e.\
\begin{equation}
\label{local}
S^{a}(w)\;V(\psi,\zeta) = \varepsilon_{\psi} V(\psi,\zeta)\; S^{a}(w)
\end{equation}
in the sense of \cite[(2.5b)]{Peter89}, where $\varepsilon_{\psi}=\pm 1$.
Using the operator product
expansion of the holomorphic field $S(w)$ with the vertex operator
$V(\psi,z)$ we can write the regular part of the meromorphic
function (\ref{mer}) as
\begin{equation}
\label{exp}
\varepsilon_{\psi}
\sum_{l=h}^{\infty} (w-z)^{l-h} \langle\varphi,V(\psi,\zeta)\;
V(S_{-l}\chi,z)\;\Omega\rangle -
\sum_{m=-\infty}^{h-1} (w-\zeta)^{m-h}
\langle\varphi, V(S_{-m}\psi,\zeta) V(\chi,z)\;\Omega\rangle;
\end{equation}
if rewritten as a power series about $w=0$ (assuming, that
(\ref{exp}) is well-defined for $w=0$, i.\ e.\ that $|\zeta|>|z|$
is suitably chosen), it
will converge for all $w$, since this function is entire. As
(\ref{mer}) is just the sum of (\ref{sing}) and (\ref{exp}),
we can use these expressions to evaluate the contour integral
(\ref{int}), thereby obtaining only terms, where $S_{l}$
either acts on $\psi$ or on $\chi$. The action is
independent of the chosen vector $\phi$. We can thus
interpret the formulae as giving a comultiplication
of the chiral algebra, namely
\begin{equation}
\label{m}
\oint_{C} dw \; w^{m}
\left\langle\phi, S^{a}(w) \;V(\psi,\zeta) V(\chi,z)\;\Omega \right\rangle
= \sum
\left\langle\phi, V(\Delta^{(1)}_{\zeta,z}(S^{a}_{m})\;\psi,\zeta)\;
V(\Delta^{(2)}_{\zeta,z}(S^{a}_{m})\;\chi,z)\;\Omega \right\rangle ,
\end{equation}
where we write $\Delta_{\zeta,z}(a)\in{\cal A}\otimes{\cal A},\;
a\in{\cal A}$ as
\begin{equation}
\Delta_{\zeta,z}(a) = \sum \Delta_{\zeta,z}^{(1)}(a)\otimes
\Delta_{\zeta,z}^{(2)}(a).
\end{equation}
We calculate
\begin{eqnarray}
{\displaystyle \Delta_{\zeta,z}(S_{n})} = &
{\displaystyle \sum_{m=1-h}^{n} \left( \begin{array}{c} n+h-1 \\ m+h-1
\end{array} \right)
\zeta^{n-m} \left(S_{m} \otimes \bbbone\right)  }
\hspace*{5.3cm} \nonumber \\
\label{chir1}
&  \hspace*{4.5cm} {\displaystyle +\, \varepsilon_{1}
\sum_{l=1-h}^{n} \left( \begin{array}{c} n+h-1 \\ l+h-1
\end{array} \right)
z^{n-l} \left(\bbbone \otimes S_{l} \right)}  \\
{\displaystyle \Delta_{\zeta,z}(S_{-n})} = &
{\displaystyle \sum_{m=1-h}^{\infty} \left( \begin{array}{c} n+m-1 \\ n-h
\end{array} \right) (-1)^{m+h-1}
\zeta^{-(n+m)} \left(S_{m} \otimes \bbbone \right) }
\hspace*{3.0cm} \nonumber \\
\label{chir2}
&  \hspace*{5.0cm} {\displaystyle +\, \varepsilon_{1}
\sum_{l=n}^{\infty} \left( \begin{array}{c} l-h \\ n-h
\end{array} \right)
(-z)^{l-n} \left(\bbbone \otimes S_{-l}  \right),}
\end{eqnarray}
where in (\ref{chir1}) we have $n\geq 1-h$, in
(\ref{chir2}) $n\geq h$ and $\varepsilon_{1}$ is $\pm 1$ according
to whether the left-hand vector in the tensor product
satisfies (\ref{local}) with $\varepsilon_{\psi}=\pm 1$.
(In (\ref{chir1}) $m$ and $l$ are in $\bbbz - h$, in
(\ref{chir2}) $m\in\bbbz -h$ and $l\in\bbbz + h$.)
\smallskip

As in \cite{MG} the above formulae are formally well defined for
$|\zeta|>1$ and $|z|<1$. If we restrict ourselves to a suitable
dense subset of the tensor product
(containing all tensor products of finite energy vectors)
the formulae are well-defined for (a suitable subset of)
$|z|<1$ and $\zeta\in\bbbc\backslash\{ 0 \}$.
We can also interchange the r\^{o}les of
$\zeta$ and $z$, thereby obtaining instead of (\ref{chir1})
\begin{eqnarray}
{\displaystyle \widetilde{\Delta}_{\zeta,z}(S_{n})} = &
{\displaystyle \varepsilon_{2}
\sum_{m=1-h}^{n} \left( \begin{array}{c} n+h-1 \\ m+h-1
\end{array} \right)
\zeta^{n-m} \left(S_{m} \otimes \bbbone\right) }
\hspace*{5.5cm} \nonumber \\
\label{chir1'}
& \hspace*{5.5cm} {\displaystyle +
\sum_{l=1-h}^{n} \left( \begin{array}{c} n+h-1 \\ l+h-1
\end{array} \right)
z^{n-l} \left(\bbbone \otimes S_{l} \right)}
\end{eqnarray}
and instead of (\ref{chir2})
\begin{eqnarray}
\label{chir2'}
{\displaystyle \widetilde{\Delta}_{\zeta,z}(S_{-n})} = &
{\displaystyle \varepsilon_{2}
\sum_{m=n}^{\infty} \left( \begin{array}{c} m-h \\ n-h
\end{array} \right)
(-\zeta)^{m-n} \left(S_{-m} \otimes \bbbone \right) }
\hspace*{2.5cm} \nonumber \\
& \hspace*{1.5cm} {\displaystyle +
\sum_{l=1-h}^{\infty} \left( \begin{array}{c} n+l-1 \\ n-h
\end{array} \right) (-1)^{l+h-1}
z^{-(n+l)} \left(\bbbone \otimes S_{l}  \right)}.
\end{eqnarray}
On the underlying conformal field theory the action of
(\ref{chir1},\,\ref{chir2}) and
(\ref{chir1'},\,\ref{chir2'}) must coincide, wherever both
are well-defined, e.\ g.\ on finite energy vectors
for $0<|z|,|\zeta| < 1$. To maintain
the whole information of conformal field theory we
have to impose these relations on the level of the
representations.  Therefore we define the ``true'' tensor
product of conformal field theory (for fixed $(\zeta,z)$) to be the
vector space tensor product quotiented by all
relations which arise in this way. On this (ring-like)
tensor product, the action of the chiral algebra
is then given by either of the two formulae.
\smallskip

We remark that we recover the formulae given in \cite{MG} for
the affine algebra and the Virasoro algebra, if we set
$h=1,2$, respectively. There we were able to prove that
(\ref{chir1} - \ref{chir2'})
define actually an algebra
homomorphism which preserves the central charge. We
cannot give a similar proof here, as we have not specified
the commutation (or anticommutation) relations of the chiral
algebra. However, recalling the derivation from conformal
field theory, it is obvious, that this has to be true
on the underlying conformal field theory. Namely,
we can use a usual contour deformation argument to see
that the action of $AB \pm BA$ on the product
of vertex operators equals the action of $[A,B]_{\pm}$, which
amounts to saying that  (\ref{chir1},\,\ref{chir2}), resp.\
(\ref{chir1'},\,\ref{chir2'}) actually
defines a comultiplication on the true tensor product
of conformal field theory.
\medskip

To proceed further we have to determine the dependence of
the tensor product on the two parameters $\zeta$ and $z$.
To this end we first translate the transformation properties
of the holomorphic field into corresponding properties
of the modes. Namely, following \cite{Peter89}, we have
\begin{equation}
e^{u L_{-1}} S(w) e^{-u L_{-1}} = S(w+u)
\end{equation}
and
\begin{equation}
\label{sc}
\lambda^{L_{0}} S(w) \lambda^{-L_{0}} = \lambda^{h} S(\lambda w) ,
\end{equation}
where in (\ref{sc}) it is understood that a choice of
the logarithm of $\lambda$ has been made. Expressing $S(w)$ in
terms of the modes (\ref{mode}) the above two equations
become
\begin{equation}
\label{trans}
e^{u L_{-1}}\; S_{m} \; e^{-u L_{-1}} = \left\{
\begin{array}{ll}
{\displaystyle \sum_{l=1-h}^{m}
\left( \begin{array}{c} m+h-1 \\ l+h-1 \end{array} \right)
(-u)^{m-l} S_{l}} & \mbox{if $m\geq 1-h$} \\
{\displaystyle \sum_{l=-m}^{\infty}
\left( \begin{array}{c} l-h \\ -m-h \end{array} \right)
u^{l+m} S_{-l}} & \mbox{if $m\leq -h$},
\end{array}
\right.
\end{equation}
and
\begin{equation}
\label{sca}
\lambda^{L_{0}} S_{m} \lambda^{-L_{0}} = \lambda^{-m} S_{m}.
\end{equation}
Using these equations we can show (as in \cite{MG})
that
\begin{equation}
\label{para}
(e^{u L_{-1}}\otimes e^{v L_{-1}})\circ\Delta_{\zeta +u,z +v}\circ
(e^{-u L_{-1}}\otimes e^{-v L_{-1}}) = \Delta_{\zeta,z}.
\end{equation}
and
\begin{equation}
\label{scal}
\left(\lambda^{L_{0}} \otimes \lambda^{L_{0}} \right)\;
\Delta_{\zeta,z}(A)\;
\left(\lambda^{-L_{0}} \otimes \lambda^{-L_{0}} \right)
= \Delta_{\lambda\zeta, \lambda z}\left(\lambda^{L_{0}} A
\lambda^{-L_{0}}\right),
\end{equation}
and similarly for $\widetilde{\Delta}$. (In the second
formula we again assume that a choice for the logarithm
of $\lambda$ has been made.)
We remark that (\ref{para}) implies that the tensor products
corresponding  to different choices of $(\zeta,z)$ are
equivalent. Thus, it does make sense to talk about
the tensor product of two representations without
reference to the insertion points, as all different
choices are equivalent.
\medskip

Next we want to investigate the associativity-properties
of the tensor product. As in \cite{MG} we can show,
that we have for suitable $\zeta_{1}, \zeta_{2}$ and $z$
\begin{equation}
\label{coass}
\left( \Delta_{\zeta_{2} - w,\zeta_{1}-w}\otimes\bbbone \right)
\circ \Delta_{w,z} =
\left( \bbbone\otimes \Delta_{\zeta_{1}-w,z-w} \right) \circ
\Delta_{\zeta_{2},w},
\end{equation}
and similarly for $\widetilde{\Delta}$ (see \cite{MG} for more
details).
These equations imply that the quotient
of the triple tensor product, we have to take out in
order to obtain the true tensor product of conformal field
theory, is independent of the bracketing.
In addition, the action of the chiral algebra is coassociative
if we choose the parameters according to (\ref{coass}). As
different choices for the parameters yield equivalent
tensor products, the tensor product
is associative up to equivalence.

We remark that using the same methods as above we could
have also determined the action of the chiral
algebra on products of three vertex operators.
A straight forward calculation shows that the
action (\ref{coass}) agrees with this action if the
vertex operators are inserted at $\zeta_{2}, \zeta_{1}$ and $z$
and we use in (\ref{exp}) the operator product
expansion of the holomorphic field with the
vertex operator at $z$. (If we take the operator product
expansion with another vertex operator, we obtain
an expression corresponding to (\ref{coass}) with some
$\Delta$'s replaced by $\widetilde{\Delta}$.)
Thus fusion of $n$ representations really corresponds to taking
successively $(n-1)$ tensor products.
\smallskip

We can similarly show that the tensor product is symmetric
up to equivalence. Namely, we can
calculate the $R$-matrix of the comultiplication, i.\ e.\
the operator $R(\zeta,z)$ in ${\cal A}\otimes {\cal A}$,
which satisfies on all tensor products
\begin{equation}
\label{rm}
R(\zeta,z)\circ\Delta_{\zeta,z}\circ R(\zeta,z)^{-1} =
\tau\circ\Delta_{\zeta,z},
\end{equation}
where $\tau$ is the
twist-map, interchanging the two factors of the tensor product.
As in \cite{MG} we find that $R(\zeta,z)$ is given by
\begin{equation}
\label{R}
R(\zeta,z)=e^{(\zeta - z) L_{-1}} \otimes e^{(z - \zeta) L_{-1}},
\end{equation}
where we assume, that $\zeta$ and $z$ are suitably chosen to ensure
convergence \cite{MG}.
\smallskip

We have thus succeeded in giving a precise definition of fusion
in the general case: fusion is simply the
tensor product introduced above with the action
of the chiral algebra induced by either of the two formulae
(\ref{chir1},\,\ref{chir2}), resp.\ (\ref{chir1'},\,\ref{chir2'}).
We have shown that the tensor product is associative
and symmetric up to equivalence.
To obtain the fusion rules we have to
study the decomposition of the tensor product into
the irreducible components. We cannot --- at the moment ---
say very much about this in the general case, however, given
any chiral algebra we can use the knowledge about null-vectors
to obtain restrictions for the possible fusion rules.

\section {$W_{3}$-algebra}
\renewcommand{\theequation}{3.\arabic{equation}}
\setcounter{equation}{0}

The $W_{3}$ algebra is generated by the Virasoro algebra
$\{ L_{n} \}$,
the modes of the energy momentum tensor with $h=2$,
and the modes $Q_{m}$ of a field of conformal
dimension $h=3$, subject to the  relations
\cite{FZ}, \cite{BW}
\begin{equation}
\label{Vira}
[ L_{n}, L_{m} ] = (n-m) L_{n+m} + \frac{1}{12}\, c\, n\, (n^{2} -1)\,
\delta_{n,-m}
\end{equation}
\begin{equation}
[ L_{m}, Q_{n} ] = (2m -n) Q_{m+n}
\end{equation}
\begin{eqnarray}
{\displaystyle [Q_{m},Q_{n}]} & = &
{\displaystyle \frac{1}{48}\, (22+5c)\, \frac{c}{3 \cdot 5!}\; (m^{2} - 4)\;
(m^{2}-1)\; m\; \delta_{m,-n} + \frac{1}{3} (m-n)\; \Lambda_{m+n}}
\nonumber \\
& & {\displaystyle
+ \frac{1}{48}\, (22+5c)\, \frac{1}{30} \;(m-n)\; (2m^{2}-mn+2n^{2}-8)\;
L_{m+n}},
\end{eqnarray}
where $\Lambda_{k}$ are the modes of a field of conformal dimension
$h_{\Lambda}=4$, which are explicitly given by
\begin{equation}
\Lambda_{n} = \sum_{k=-\infty}^{\infty} : L_{n-k} L_{k} :
+ \frac{1}{5} x_{n} L_{n},
\end{equation}
$x_{2l}=(l+1)(1-l)$ and $x_{2l+1}=(l+2)(1-l)$.
Evaluating the general comultiplication formulae
(\ref{chir1},\,\ref{chir2}) for the present
case we obtain
\begin{equation}
\label{vir1}
{\displaystyle \Delta_{\zeta,z}(L_{n})} =
{\displaystyle \sum_{m=-1}^{n} \left( \begin{array}{c} n+1 \\ m+1
\end{array} \right)
\zeta^{n-m} \left(L_{m} \otimes \bbbone \right) +
\sum_{l=-1}^{n} \left( \begin{array}{c} n+1 \\ l+1
\end{array} \right)
z^{n-l} \left(\bbbone \otimes L_{l} \right)} \hspace*{0.4cm}
\end{equation}
\begin{eqnarray}
{\displaystyle \Delta_{\zeta,z}(L_{-n})} = &
{\displaystyle \sum_{m=-1}^{\infty} \left( \begin{array}{c} n+m-1 \\ n-2
\end{array} \right) (-1)^{m+1}
\zeta^{-(n+m)} \left(L_{m} \otimes \bbbone \right)} \hspace*{3.5cm}
\nonumber \\
\label{vir2}
& \hspace*{5.5cm} {\displaystyle +
\sum_{l=n}^{\infty} \left( \begin{array}{c} l-2 \\ n-2
\end{array} \right)
(-z)^{l-n} \left(\bbbone \otimes L_{-l} \right)},
\end{eqnarray}
and
\begin{eqnarray}
\label{w31}
{\displaystyle \Delta_{\zeta,z}(Q_{n})} = &
{\displaystyle \sum_{m=-2}^{n} \left( \begin{array}{c} n+2 \\ m+2
\end{array} \right)
\zeta^{n-m} \left(Q_{m} \otimes \bbbone\right) +
\sum_{l=-2}^{n} \left( \begin{array}{c} n+2 \\ l+2
\end{array} \right)
z^{n-l} \left(\bbbone \otimes Q_{l} \right)}  \\
{\displaystyle \Delta_{\zeta,z}(Q_{-n})} = &
{\displaystyle \sum_{m=-2}^{\infty} \left( \begin{array}{c} n+m-1 \\ n-3
\end{array} \right) (-1)^{m}
\zeta^{-(n+m)} \left(Q_{m} \otimes \bbbone \right) }
\hspace*{3.5cm} \nonumber \\
\label{w32}
& \hspace*{5.5cm} {\displaystyle +
\sum_{l=n}^{\infty} \left( \begin{array}{c} l-3 \\ n-3
\end{array} \right)
(-z)^{l-n} \left(\bbbone \otimes Q_{-l}  \right)},
\end{eqnarray}
where in (\ref{vir1}) we have $n\geq -1$, in (\ref{vir2})
$n\geq 2$, in (\ref{w31}) $n\geq -2$ and in
(\ref{w32}) $n\geq 3$.
\smallskip

The comultiplication of $\Lambda_{n}$
is determined by the formulae (\ref{chir1}) and (\ref{chir2}),
since the $\Lambda_{n}$'s are the modes of a field
of conformal dimension $h=4$. We can thus show,
following the methods of \cite{MG} and
using the power series expansions of appendix A,
that these formulae do indeed define a comultiplication.
\smallskip

The highest weight representations are labeled by the eigenvalues
of $L_{0}$ and $Q_{0}$ of the highest weight vector $|h,q>$, i.\ e.\
\begin{equation}
L_{m} |h,q> = \delta_{m,0}\;h\; |h,q>, \hspace{0.5cm}
Q_{m} |h,q> = \delta_{m,0}\;q\; |h,q>, \hspace{1.0cm}
m\geq 0.
\end{equation}
We parametrize the weights of a $W$-highest weight vector following
\cite{FZ}, \cite{BW} as
\begin{equation}
h=\frac{1}{3}\, (x^{2} + xy + y^{2} - 3 a^{2}) \hspace{0.5cm}
q=\frac{1}{27}\, (x-y)\; (2x+y)\; (x+2y),
\end{equation}
\begin{equation}
x= p \alpha - q / \alpha \hspace{0.5cm}
y= r \alpha - s / \alpha,
\end{equation}
where we define $a$ and $\alpha$ by
\begin{equation}
c=2 -24 a^{2} \hspace{0.5cm}
\alpha= \frac{a}{2} \pm \sqrt{\frac{a^{2}}{4} +1}.
\end{equation}
It is believed \cite{FZ} \cite{BW} that the doubly degenerate primary fields,
i.\ e.\ those which have two independent null-vectors, satisfy
$p,q,r,s\in\bbbn$. In particular the representation $(1,1;1,1)$ is
just the vacuum.
\medskip

To derive restrictions on the possible fusion rules for
representations of the $W_{3}$ algebra, we use the
explicit form of the null-vectors of the highest
weight vectors $\phi_{(p,r;q,s)}$ corresponding to
$(p,r;q,s)=(2,1;1,1),\;(1,2;1,1),\;(1,1;2,1),\;(1,1;1,2)$,
given by
\begin{equation}
\left( 2\, h_{(p,r;q,s)}\; Q_{-1} - 3\,q_{(p,r;q,s)}\; L_{-1} \right)
\phi_{(p,r;q,s)}
\end{equation}
and
\begin{equation}
\left( (5 h_{(p,r;q,s)} +1)\, h_{(p,r;q,s)}\; Q_{-2} - 12
q_{(p,r;q,s)}\; L_{-1}^{2} + 6 q_{(p,r;q,s)}\,
(h_{(p,r;q,s)} +1)\; L_{-2} \right) \phi_{(p,r;q,s)}.
\end{equation}
Suppose that the representation $(p',r';q',s')$ is
contained in the tensor product of $(2,1;1,1)$ and
$(p,r;q,s)$ corresponding to $(\zeta,z)=(\zeta,0)$. (As
the tensor product is equivalent for all different choices
of the parameters $(\zeta,z)$,
we can restrict ourselves without loss of generality
to this case.)
Then the scalar product (in the following we omit
the subscripts of the tensor product and the
intertwiner $\pi_{\zeta,0}$, cf. \cite{MG})
\begin{equation}
\left\langle \phi_{(p',r';q',s')}, \left(\phi_{(p,r;q,s)}\otimes
\phi_{(2,1;1,1)} \right) \right\rangle
\end{equation}
does not vanish.
As $\phi_{(p',r';q',s')}$ is a highest weight vector, we
have the two equations (we write $\phi'=\phi_{(p',r';q',s')},\;
\phi=\phi_{(p,r;q,s)}$ and $\phi_{0}=\phi_{(2,1;1,1)}$ and
similarly for $h$ and $q$)
\begin{equation}
\label{equai}
0=\left\langle \phi',
\left( 2 h_{0}\; \Delta\,(Q_{-1})
- 3\,q_{0}\;\Delta\,(L_{-1}) \right)
\left(\phi\otimes \phi_{0}\right) \right\rangle
\end{equation}
and
\begin{equation}
\label{equaii}
0=\left\langle \phi',
\left( (5 h_{0}+1)\;h_{0}\;\Delta\,(Q_{-2})
- 12 q_{0}\; \Delta\,(L_{-1})^{2}
+ 6 q_{0}\,(h_{0} +1)\; \Delta\,(L_{-2})
\right)
\left(\phi\otimes \phi_{0}\right) \right\rangle.
\end{equation}
\smallskip

We can use these equations to obtain necessary restrictions
for the fusion rules of the $W_{3}$-algebra, performing an
analysis similar to \cite{MG} (see appendix B for details).
We arrive at
\footnote{These conditions have also been
independently derived in \cite{BPT} and \cite{BW1}.}
\begin{equation}
\label{w32111}
\phi_{(2,1;1,1)} \otimes \phi_{(p,q;r,s)} =
\phi_{(p+1,q;r,s)} \oplus \phi_{(p,q-1;r,s)}
\oplus \phi_{(p-1,q+1;r,s)}
\end{equation}
\begin{equation}
\label{w31211}
\phi_{(1,2;1,1)} \otimes \phi_{(p,q;r,s)} =
\phi_{(p,q+1;r,s)} \oplus \phi_{(p-1,q;r,s)}
\oplus \phi_{(p+1,q-1;r,s)}
\end{equation}
\begin{equation}
\label{w31121}
\phi_{(1,1;2,1)} \otimes \phi_{(p,q;r,s)} =
\phi_{(p,q;r+1,s)} \oplus \phi_{(p,q;r,s-1)}
\oplus \phi_{(p,q;r-1,s+1)}
\end{equation}
\begin{equation}
\label{w31112}
\phi_{(1,1;1,2)} \otimes \phi_{(p,q;r,s)} =
\phi_{(p,q;r,s+1)} \oplus \phi_{(p,q;r-1,s)}
\oplus \phi_{(p,q;r+1,s-1)} .
\end{equation}
Assuming that the representation
$\phi_{(1,2;1,1)}$ is indeed contained in the tensor product
of $\phi_{(2,1;1,1)}\otimes\phi_{(2,1;1,1)}$ and
similarly for $\phi_{(2,1;1,1)}$ etc. --- we should be
able to prove this using the methods of \cite{BFIZ}, \cite{BW1} ---
we can already derive restrictions for the general
fusion rules from these equations.
Firstly we observe, that we might restrict ourselves
to the set of representations, which
are contained in tensor products of the form
\begin{equation}
\label{tp}
\phi_{(2,1;1,1)} ^{p} \otimes \phi_{(1,2;1,1)} ^{q} \otimes
\phi_{(1,1;2,1)} ^{r} \otimes \phi_{(1,1;1,2)} ^{s} \otimes
\phi_{(1,1;1,1)},
\end{equation}
where $p,q,r,s\in\bbbn_{0}$. If $\phi_{(p_{0},q_{0};r_{0},s_{0})}$
is contained in this set of representations,
then it is automatically contained in the tensor
product (\ref{tp}) with $p_{0}=p-1,\; q_{0}=q-1$ etc. --- this
can be seen using an induction argument.
Next we can show by induction on $p$,
that for $p\geq 2$
\begin{equation}
\phi_{(2,1;1,1)} \otimes \phi_{(p,1;1,1)}
=\phi_{(p+1,1;1,1)} \oplus \phi_{(p-1,2;1,1)}
\hspace{1.5cm}
\phi_{(1,2;1,1)} \otimes \phi_{(p,1;1,1)}
=\phi_{(p,2;1,1)} \oplus \phi_{(p-1,1;1,1)}
\end{equation}
and similarly for $(1,q;1,1)$, etc.
This establishes that the set of all representations
$\phi_{(p,q;r,s)}$ for $p,q,r,s\in\bbbn$
is closed under the operation of taking tensor products.
Furthermore, using the associativity and
symmetry of the tensor product as in \cite{MG}, we can
derive restrictions for the general fusion rules, namely,
\begin{equation}
\phi_{(p_{2},q_{2};r_{2},s_{2})}\otimes
\phi_{(p_{1},q_{1};r_{1},s_{1})}
= \sum_{(p,q)\in\Lambda(p_{1},q_{1},p_{2},q_{2}) } \;
  \sum_{(r,s)\in\Lambda(r_{1},s_{1},r_{2},s_{2}) }
  \left[ \phi_{(p,q;r,s)} \right] ,
\end{equation}
where $\Lambda(p_{1},q_{1},p_{2},q_{2})$ consists of all
pairs $(p,q)\in\bbbn\times\bbbn$, which satisfy
\begin{equation}
(p,q)=(p_{1},q_{1}) + \lambda_{2} = (p_{2},q_{2}) + \lambda_{1}
\end{equation}
for some $\lambda_{i}\in\Lambda[p_{i}-1,q_{i}-1]$, where
$\Lambda[l,m]$ denotes the weight lattice of the irreducible
representation of $A_{2}$ corresponding to the weight
$[lm]$ in the Dynkin-basis.

So far we have only analyzed the case, where $c$ is generic.
For special values of $c$, corresponding to the minimal
models, we have additional null-vectors, which
give rise to a truncation of the fusion rules. We then obtain
a finite set of representations closed
under the operation of taking tensor products.
We remark that the above restrictions have already been derived
in \cite{FZ} by means of a free field construction and in
\cite{FKW} via quantized Drinfeld-Sokolov reduction.

\section{The $N=1$ NS superconformal algebra}
\renewcommand{\theequation}{4.\arabic{equation}}
\setcounter{equation}{0}

The $N=1$ NS superconformal algebra is generated by the
Virasoro algebra (\ref{Vira})  $\{L_{n}\},\, n\in\bbbz$, and the
modes of the superfield $G$ with $h=\thalf$,
$\{ G_{\alpha}\}, \alpha\in\bbbz + \half$,
subject to the relations
\begin{equation}
[ L_{n}, G_{\alpha} ] = \left( \half n - \alpha \right) G_{n+\alpha}
\end{equation}
\begin{equation}
\{ G_{\alpha}, G_{\beta} \} = 2 L_{\alpha+\beta} + \frac{1}{3} c
\left( \alpha^{2} - \frac{1}{4} \right) \delta_{\alpha, -\beta}.
\end{equation}
Evaluating (\ref{chir1},\,\ref{chir2}) we obtain for the
Virasoro algebra (\ref{vir1},\,\ref{vir2}) and for
the $G_{\mu}$'s
\begin{eqnarray}
\label{g1}
{\displaystyle \Delta_{\zeta,z}(G_{\alpha})} = &
{\displaystyle \sum_{\mu=-\half}^{\alpha} \left(
\begin{array}{c} \alpha + \half \\ \mu + \half
\end{array} \right)
\zeta^{\alpha-\mu} \left(G_{\mu} \otimes \bbbone\right) + \varepsilon_{1}
\sum_{\lambda=-\half}^{\alpha} \left(
\begin{array}{c} \alpha + \half \\ \lambda + \half
\end{array} \right)
z^{\alpha-\lambda} \left(\bbbone \otimes G_{\lambda} \right)}  \\
{\displaystyle \Delta_{\zeta,z}(G_{-\alpha})} = &
{\displaystyle \sum_{\mu=-\half}^{\infty} \left(
\begin{array}{c} \alpha+\mu-1 \\ \alpha -\thalf
\end{array} \right) (-1)^{\mu+\half}
\zeta^{-(\alpha+\mu)} \left(G_{\mu} \otimes \bbbone \right) }
\hspace*{3.5cm} \nonumber \\
\label{g2}
& \hspace*{5.0cm} {\displaystyle +\, \varepsilon_{1}
\sum_{\lambda=\alpha}^{\infty} \left(
\begin{array}{c} \lambda-\thalf \\ \alpha - \thalf
\end{array} \right)
(-z)^{\lambda-\alpha} \left(\bbbone \otimes G_{-\lambda}  \right)},
\end{eqnarray}
where in (\ref{g1}) $\alpha\geq - \half$ and in
(\ref{g2}) $\alpha\geq \thalf$.
\smallskip

We remark that $\varepsilon_{G_{\alpha}\psi}= - \varepsilon_{\psi}$ and
$\varepsilon_{L_{n}\psi}=\varepsilon_{\psi}$. It is then easy to
check, using the power series expansions of appendix A,
that the above formulae do indeed define a comultiplication.
\medskip

Following \cite{E} we consider the degenerate representations,
parametrized by the eigenvalue of $L_{0}$
\begin{equation}
\label{hdelta}
h_{m,n}= h_{0} + \left( \half \alpha_{+} m +
\half \alpha_{-} n \right)^{2} ,
\end{equation}
where $m,n\in\half\bbbn,\; m-n\in\bbbz$, and
\begin{equation}
h_{0}= \frac{1}{16} \left(\frac{2}{3} \, c - 1\right), \hspace{1.5cm}
\alpha_{\pm} = \half \left[
\sqrt{1-\frac{2}{3} \, c} \pm \sqrt{9-\frac{2}{3} c} \right].
\end{equation}
In this parametrization the vacuum representation is just
$(\half,\half)$.

To derive restrictions for the fusion rules we shall
use the explicit form of the first few singular
vectors, namely
\begin{equation}
\label{super13}
\left( G_{-\thalf} - \frac{2}{(2 h_{m,n} + 1 )}\; G_{-\half}\;
L_{-1} \right) \phi_{m,n}
\end{equation}
where $(m,n)=(\half,\thalf)$ or $(m,n)=(\thalf,\half)$, and
\begin{equation}
\label{super11}
\left( L_{-1}^{2} - \frac{4}{3} h_{1,1}\; L_{-2}
- G_{-\thalf}\;G_{-\half} \right) \phi_{1,1}.
\end{equation}
Firstly, we analyze the tensor product of $\phi_{\half,\thalf}$
with $\phi_{m,n}$. Suppose the representation generated by
the highest weight vector
$\phi'$ is contained in the tensor product and
the scalar product
\begin{equation}
\label{scaln1}
\left\langle \phi', (\phi_{m,n} \otimes \phi_{\half,\thalf})
\right\rangle
\end{equation}
does not vanish. As $\phi'$ is
a highest weight vector, we have (setting $z=0$ and
writing $\phi=\phi_{m,n},\;\phi_{0}=\phi_{\half,\thalf}$
and similarly for $h$)
\begin{eqnarray}
0 & = &
{\displaystyle
\left\langle \phi',
\Delta\,( G_{\half})\;
\left(\Delta\,(G_{-\thalf}) - \frac{2}{(2 h_{0} + 1 )}
\Delta\,(G_{-\half}) \;  \Delta\,(L_{-1}) \right)
(\phi \otimes \phi_{0})
\right\rangle} \nonumber \\
& = &
{\displaystyle
\left\langle \phi',
(L_{-1}\otimes\bbbone)\; (\phi \otimes \phi_{0} )
\right\rangle
+ 2\; h\; \zeta^{-1}
\left\langle \phi', (\phi \otimes \phi_{0}) \right\rangle} \nonumber \\
& & {\displaystyle
- \frac{2}{(2 h_{0} + 1 )}\, \left[\;
\left\langle \phi', \Delta\,(G_{\half})\; \Delta\,(L_{-1})\;
\left( G_{-\half}\otimes\bbbone\right)
(\phi \otimes \phi_{0}) \right\rangle \right. } \nonumber \\
& & {\displaystyle
+ \, \varepsilon_{1} \, \zeta
\left\langle \phi',
(G_{-\half}\;L_{-1} \otimes G_{-\half})\;
(\phi \otimes \phi_{0}) \right\rangle
+ \varepsilon_{1} \, \left\langle \phi',
(G_{\half}\;L_{-1} \otimes G_{-\half})\;
(\phi \otimes \phi_{0}) \right\rangle } \nonumber \\
& & {\displaystyle \left. + \left\langle \phi',
(L_{-1} \otimes G_{\half}\;G_{-\half})\;
(\phi \otimes \phi_{0}) \right\rangle
\right] },
\end{eqnarray}
where we have used (\ref{super13}).
Using the identities
\begin{eqnarray}
{\displaystyle
\left\langle \phi',
(G_{-\half}\; L_{-1} \otimes G_{-\half})\;
(\phi \otimes \phi_{0}) \right\rangle } & = &
{\displaystyle
\left\langle \phi',
\Delta\,(G_{-\half})\;(L_{-1}\otimes G_{-\half})\;
(\phi \otimes \phi_{0}) \right\rangle } \nonumber \\
& &
{\displaystyle - \varepsilon_{1}
\left\langle \phi',
(L_{-1}\otimes G_{-\half}^{2})\;
(\phi \otimes \phi_{0}) \right\rangle } \nonumber \\
& = &
{\displaystyle - \varepsilon_{1}
\left\langle \phi',
(L_{-1}\otimes L_{-1})\;
(\phi \otimes \phi_{0}) \right\rangle}
\end{eqnarray}
and
\begin{eqnarray}
{\displaystyle
\varepsilon_{1}\, \left\langle \phi',
(G_{\half}\; L_{-1} \otimes G_{-\half})\;
(\phi \otimes \phi_{0}) \right\rangle } & = &
{\displaystyle \varepsilon_{1}\,
\left\langle \phi',
(G_{-\half}\otimes G_{-\half})\;
(\phi \otimes \phi_{0}) \right\rangle } \nonumber \\
& = &
{\displaystyle
- \left\langle \phi',
\Delta\,(G_{-\half})\;(G_{-\half}\otimes \bbbone)\;
(\phi \otimes \phi_{0}) \right\rangle } \nonumber \\
& & {\displaystyle
+ \left\langle \phi',
(L_{-1} \otimes \bbbone)\;
(\phi \otimes \phi_{0}) \right\rangle }
\end{eqnarray}
together with (\ref{kappa1},\,\ref{kappa2}) we obtain
\begin{eqnarray}
0  & = &
{\displaystyle
\zeta^{-1} \left[\kappa + 2\; h - \frac{2}{(2 h_{0} + 1 )}
\left(\kappa \;(\kappa-1) + \kappa \;(2\;h_{0} +1) \right) \right]
\left\langle \phi', \left(\phi \otimes \phi_{0}\right) \right\rangle}
\nonumber \\
& = &
{\displaystyle
\zeta^{-1} \left[ - \kappa + 2\; h - \frac{2}{(2 h_{0} + 1 )}\;
\kappa\;(\kappa - 1) \right]
\left\langle \phi', \left(\phi \otimes \phi_{0}\right) \right\rangle},
\end{eqnarray}
where $\kappa = h' - h - h_{0}$. Again this formula
implies $[..]=0$, which is a quadratic equation in $h'$. For each of
the two fields $\phi_{\half,\thalf}$ and $\phi_{\thalf,\half}$
we have thus two solutions, giving the fusion rules \cite{E}
\begin{equation}
\phi_{m,n} \otimes \phi_{\half,\thalf} =
\phi_{m,n+1} \oplus \phi_{m,n-1}
\end{equation}
and
\begin{equation}
\phi_{m,n} \otimes \phi_{\thalf,\half} =
\phi_{m+1,n} \oplus \phi_{m-1,n}.
\end{equation}
Similarly, we can use
\begin{equation}
0=\left\langle \phi',
\left( \Delta\,(L_{-1})^{2} - \frac{4}{3}\, h_{0}\,
\Delta\,(L_{-2}) - \Delta\,(G_{-\thalf})\;
\Delta\,(G_{-\half}) \right)
(\phi \otimes \phi_{0}) \right\rangle
\end{equation}
to derive restrictions for the possible fusions involving
the field $\phi_{1,1}$.
With the help of the identity
\begin{eqnarray}
{\displaystyle
\left\langle \phi',
\left(G_{-\half}\otimes G_{-\thalf}\right)\;
(\phi \otimes \phi_{0}) \right\rangle}
& = &
{\displaystyle
- \varepsilon_{1} \left\langle \phi',
\Delta\,(G_{-\thalf})\;\left(G_{-\half}\otimes\bbbone\right)\;
(\phi \otimes \phi_{0}) \right\rangle } \nonumber \\
& & {\displaystyle
+ \varepsilon_{1}\; \zeta^{-1} \;
\left\langle \phi',
(L_{-1}\otimes\bbbone)\;
(\phi \otimes \phi_{0}) \right\rangle} \nonumber \\
& &
{\displaystyle -\varepsilon_{1}\;\zeta^{-2}\;
\left\langle \phi',
(G_{\half}\, G_{-\half}\otimes\bbbone)\;
(\phi \otimes \phi_{0}) \right\rangle} \nonumber \\
& = &
{\displaystyle \zeta^{-2}\; \varepsilon_{1}
\left( \kappa - 2 h \right)
\left\langle \phi',
(\phi \otimes \phi_{0}) \right\rangle}
\end{eqnarray}
we obtain
\begin{equation}
0= \zeta^{-2}
\left[- \kappa^{2} - \frac{4}{3}\, h_{0} \;(\kappa - h) \right]
\left\langle \phi', \left(\phi \otimes \phi_{0}\right) \right\rangle,
\end{equation}
which implies \cite{E}
\begin{equation}
\phi_{m,n} \otimes \phi_{1,1} =
\phi_{m-\half, n-\half} \oplus
\phi_{m+\half,n+\half}.
\end{equation}
Restricting ourselves to representations, which are
contained in
\begin{equation}
\label{tpn1}
\phi_{\thalf,\half}^{p} \otimes \phi_{\half,\thalf}^{q}
\otimes \phi_{1,1}^{\eta} \otimes \phi_{\half,\half},
\end{equation}
where $p,q\in\bbbn_{0}$ and $\eta=0,1$, we can again
show, that if a representation is contained in this set
then it is contained in the tensor product (\ref{tpn1}) with
the minimal values for $p$ and $q$.
As in \cite{MG} we can thus
derive restrictions for the fusion rules in the general case, using
the associativity and symmetry of the tensor product.
Depending on whether $(\alpha,\beta)$ and $(\alpha',\beta')$ are
half-integers or integers, we obtain
\begin{equation}
\label{case}
\phi_{\alpha,\beta} \otimes \phi_{\alpha',\beta'} =
\left\{
\begin{array}{lr}
{\displaystyle \sum_{\gamma= |\alpha-\alpha'| + \half}^{\alpha+\alpha'-\half}\;
\sum_{\delta= |\beta-\beta'| + \half}^{\beta+\beta'-\half}\;
\left[ \phi_{\gamma,\delta} \right]}
& \mbox{if $\alpha,\beta,\alpha',\beta' \in\bbbz + \half$ }
\vspace*{0.3cm} \\

{\displaystyle
\sum_{\gamma= |\alpha -\alpha'- \half| + 1}^{\alpha+\alpha'-\half}\;
\sum_{\delta= |\beta - \beta' - \half| + 1}^{\beta+\beta' -\half}\;
\left[ \phi_{\gamma,\delta} \right]} \hspace*{0.5cm}
& \mbox{if $\alpha,\beta\in \bbbz,\; \alpha', \beta' \in \bbbz +
\half$ } \vspace*{0.3cm} \\

{\displaystyle
\sum_{\gamma= |\alpha-\alpha'| +\half}^{\alpha+\alpha'-\half}\;
\sum_{\delta= |\beta - \beta'| + \half}^{\beta+\beta' - \half}\;
\left[ \phi_{\gamma,\delta} \right]}
& \mbox{if $\alpha,\beta,\alpha',\beta' \in\bbbz$ , }
\end{array}
\right.
\end{equation}
where in the first two lines of (\ref{case})
$\gamma$ and $\delta$ attain only every other value and
in the third line of (\ref{case}) the summands corresponding
to $\gamma=\alpha + \alpha' - \half, \delta=|\beta - \beta'|
+ \half$ and $ \gamma=|\alpha - \alpha'| + \half, \;
\delta=\beta+\beta'-\half$ are excluded.
The above are exactly the even fusion rules of \cite{SS}.

In the $W_{3}$-case we could express the scalar products
of the form $\left\langle \phi', \left( Q_{-p}\otimes\bbbone\right)
\left(\phi\otimes\phi_{0}\right) \right\rangle$ with
$p=1,2$ in terms of $\left\langle \phi',
\left(\phi\otimes\phi_{0}\right) \right\rangle$.
In the present case, however, we cannot obtain such a relation,
i.\ e.\  we cannot express
\begin{equation}
\label{scaln1'}
\left\langle \phi',
\left(G_{-\half}\otimes \bbbone \right)
(\phi_{m,n} \otimes \phi_{\half,\thalf})
\right\rangle =
- \varepsilon_{1}\;
\left\langle \phi',
\left(\bbbone\otimes G_{-\half} \right)
(\phi_{m,n} \otimes \phi_{\half,\thalf})
\right\rangle
\end{equation}
in terms of $\left\langle \phi', (\phi_{m,n} \otimes \phi_{\half,\thalf})
\right\rangle$. \footnote{I thank G. Watts for pointing this out
to me.}
Thus there exist highest weight vectors $\phi'$
such that (\ref{scaln1}) is zero, but not (\ref{scaln1'}).
In this case the scalar product of the three highest weight
vectors vanishes, but the representation generated from $\phi'$
is nevertheless contained in the tensor product.

We remark that the $(\zeta,z)-$dependence of the scalar product
(\ref{scaln1'}) is given by
\begin{equation}
(\zeta - z)^{h'-h-h_{0} + \half},
\end{equation}
which follows from the same reasoning as in \cite{MG}. As
the scalar product corresponds essentially to the three-point-function,
we expect that this case corresponds to the odd fusion rules
of \cite{SS}. Indeed, carrying through the analogous analysis,
we can check that we obtain in this way the same restrictions as
for the odd fusion rules \cite{SS}.

\section{The $N=2$ NS superconformal algebra}
\renewcommand{\theequation}{5.\arabic{equation}}
\setcounter{equation}{0}

The $N=2$ NS superconformal algebra is generated by the Virasoro
algebra (\ref{Vira}), the modes of an $U(1)$-current with $h=1$
$\{T_{n}\},\,n\in\bbbz$ and the modes of two superfields of
conformal dimension $h=\thalf$, $\{G^{\pm}_{\alpha}\},\,
\alpha\in\bbbz+\half$, subject to the relations \cite{BFK}
\begin{equation}
\begin{array}{ccl}
{\displaystyle
\left[ L_{m}, G^{\pm}_{\alpha}\right]} & = &
{\displaystyle
\left(\half m - \alpha \right)\; G^{\pm}_{\alpha+m} } \nonumber \\
\vspace{0.2cm}

{\displaystyle
\left[ L_{m}, T_{n} \right] } & = &
{\displaystyle
- n \; T_{m+n}} \nonumber \\
\vspace{0.2cm}

{\displaystyle
\left[ T_{m},T_{n} \right]} & = &
{\displaystyle \ct\; m \;\delta_{m,-n}} \nonumber \\
\vspace{0.2cm}

{\displaystyle
\left[ T_{m}, G^{\pm}_{\alpha} \right] } & = &
{\displaystyle \pm G^{\pm}_{\alpha+m} } \nonumber \\
\vspace{0.2cm}

{\displaystyle
\left\{ G^{\pm}_{\alpha}, G^{\pm}_{\beta} \right\}} & = &
0 \nonumber \\
{\displaystyle
\left\{ G^{+}_{\alpha}, G^{-}_{\beta} \right\}} & = &
{\displaystyle
2 \;  L_{\alpha + \beta} + (\alpha - \beta) \; T_{\alpha + \beta}
+ \ct \;\left( \alpha^{2} - \frac{1}{4} \right) \delta_{\alpha,-\beta}, }
\end{array}
\end{equation}
where $\ct = c/3$.
The comultiplication is given by (\ref{vir1},\,\ref{vir2}) for
the Virasoro algebra, (\ref{g1},\,\ref{g2}) for the superfield modes
and by
\begin{equation}
\label{u1}
{\displaystyle \Delta_{\zeta,z}(T_{n})} =
{\displaystyle \sum_{m=0}^{n} \left( \begin{array}{c} n \\ m
\end{array} \right)
\zeta^{n-m} \left(T_{m} \otimes \bbbone \right) +
\sum_{l=0}^{n} \left( \begin{array}{c} n \\ l
\end{array} \right)
z^{n-l} \left(\bbbone \otimes T_{l} \right)} \hspace*{1.4cm}
\end{equation}
\begin{eqnarray}
{\displaystyle \Delta_{\zeta,z}(T_{-n})} = &
{\displaystyle \sum_{m=0}^{\infty} \left( \begin{array}{c} n+m-1 \\ n-1
\end{array} \right) (-1)^{m}
\zeta^{-(n+m)} \left(T_{m} \otimes \bbbone \right) } \hspace*{3.5cm}
\nonumber \\
\label{u2}
& \hspace*{5.5cm} {\displaystyle +
\sum_{l=n}^{\infty} \left( \begin{array}{c} l-1 \\ n-1
\end{array} \right)
(-z)^{l-n} \left(\bbbone \otimes T_{-l} \right)},
\end{eqnarray}
where in (\ref{u1}) $n\geq 0$ and in (\ref{u2}) $n\geq 1$.

We restrict ourselves for simplicity to the
doubly-degenerate representations
which have two fermionic null-vectors. These representations
contain in particular all unitary representations with
$c<1$ \cite{BFK}.
We can parametrize these representations
by two half-odd integers $j,k \in \bbbz + \half,\,
j,k>0$,
where the representation $(j,k)$ has null-vectors at
$(\Delta h, \Delta q)= (j,1)$ and $(\Delta h, \Delta q)= (k,-1)$ ---
$\Delta h$ denotes the relative $L_{0}-$eigenvalue and
$\Delta q$ the relative $T_{0}-$eigenvalue.
The conformal weight and charge of the corresponding highest
weight vector is then given by
\begin{equation}
q=\half \left( \ct -1 \right)\;\left( j-k \right) \hspace*{1.5cm}
h=\half \left( \ct -1 \right)\;\left( \frac{1}{4} - j\, k\right).
\end{equation}
We remark that the representation $(\half,\half)$ is just the vacuum
representation. Furthermore, we shall use the explicit form of
the two null-vectors of the representations $(\half, \thalf)$
\begin{equation}
G^{+}_{-\half} \phi_{\half,\thalf}
\end{equation}
and
\begin{equation}
\left( (q_{\half,\thalf} + \ct - 1)\; G^{-}_{-\thalf} +
L_{-1}\; G^{-}_{-\half} + T_{-1}\; G^{-}_{-\half} \right)
\phi_{\half,\thalf}
\end{equation}
and of the representation $(\thalf,\half)$
\begin{equation}
G^{-}_{-\half} \phi_{\thalf,\half}
\end{equation}
and
\begin{equation}
\left( (\ct - 1 - q_{\thalf,\half} )\; G^{+}_{-\thalf} +
L_{-1}\; G^{+}_{-\half} - T_{-1}\; G^{+}_{-\half} \right)
\phi_{\thalf,\half}.
\end{equation}
\medskip

To obtain restrictions for the even fusion rules let us first
analyze the tensor product of $\phi_{j,k}$ with
$\phi_{\half,\thalf}$. Suppose the irreducible
representation  generated by the
highest weight vector $\phi'$ is contained in the tensor product,
then the scalar product
\begin{equation}
\label{scaln2}
\left\langle \phi', \left(\phi_{j,k} \otimes \phi_{\half,\thalf}\right)
\right\rangle
\end{equation}
does not vanish. On the other hand, as $\phi'$ is a highest
weight vector, we have (setting $z=0$ and writing $\phi=\phi_{j,k},\;
\phi_{0}=\phi_{\half,\thalf}$ and similarly for $h$ and $q$)
\begin{equation}
\left( 2 h' - q' \right)\;
\left\langle \phi', \left(\phi \otimes \phi_{0}\right) \right\rangle =
\left\langle \phi',
\Delta\, (G^{-}_{\half})\; \Delta \, (G^{+}_{-\half})\;
\left( \phi \otimes \phi_{0}\right) \right\rangle ,
\end{equation}
from which we can conclude that
\begin{equation}
\label{equa1}
\left\langle \phi',
\left( G^{-}_{-\half}\; G^{+}_{-\half} \otimes \bbbone \right)
\left(\phi \otimes \phi_{0}\right) \right\rangle =
\zeta^{-1}\; \left(2 (h' -h) + q - q' \right)
\left\langle \phi', \left(\phi \otimes \phi_{0}\right) \right\rangle.
\end{equation}
Similarly, we can use the other null-vector equation to obtain
\begin{eqnarray}
0 & = &
{\displaystyle \left\langle \phi',
\Delta\,(G^{+}_{\half}) \left[
\mu\, \Delta\, (G^{-}_{-\thalf}) + \Delta\, (L_{-1}) \;
\Delta\, (G^{-}_{-\half}) + \Delta\, (T_{-1})\;
\Delta\, (G^{-}_{-\half}) \right]
\left(\phi \otimes \phi_{0}\right) \right\rangle} \nonumber \\
& = &
{\displaystyle \left\langle \phi',
\left[ \Delta\,(G^{+}_{\half})\; \Delta\,(L_{-1})
+ \Delta\,(G^{+}_{\half})\; \Delta\,(T_{-1}) \right]
\left( G^{-}_{-\half}\otimes\bbbone\right)
\left(\phi \otimes \phi_{0}\right) \right\rangle}
\nonumber \\
& &
{\displaystyle + \varepsilon_{1}
\left\langle \phi',
\Delta\,(G^{+}_{\half})
\left[ \left( L_{-1} \otimes G^{-}_{-\half}\right)
+ q\, \zeta^{-1}\,
\left( \bbbone \otimes G^{-}_{-\half}\right) \right]
\left(\phi \otimes \phi_{0}\right) \right\rangle} \nonumber \\
& &
{\displaystyle + \mu\, \zeta^{-1}\left\langle \phi',
\Delta\,(G^{+}_{\half}) \left( G^{-}_{-\half}\otimes\bbbone\right)
\left(\phi \otimes \phi_{0}\right) \right\rangle} \nonumber \\
& = &
{\displaystyle
\varepsilon_{1} \left( \kappa + q \right)
\left\langle \phi',
\left( G^{+}_{-\half} \otimes G^{-}_{-\half}\right)
\left(\phi \otimes \phi_{0}\right) \right\rangle
+ \mu
\left\langle \phi',
\left( G^{+}_{-\half} \; G^{-}_{-\half}\otimes\bbbone\right)
\left(\phi \otimes \phi_{0}\right) \right\rangle} \nonumber \\
& &
{\displaystyle
+ \left(\mu \left(2\,h + q \right) + q \left( 2\, h_{0} + q_{0}\right)
+ \kappa \left(2\, h_{0} + q_{0} \right) \right) \zeta^{-1}
\left\langle \phi',\left(\phi \otimes \phi_{0}\right) \right\rangle,}
\end{eqnarray}
where $\mu= q_{0} + \ct - 1$ and $\kappa = h' -h - h_{0}$.
Using the identity
\begin{eqnarray}
0 & = &
{\displaystyle
\left\langle \phi', \Delta\,(G^{-}_{-\half}) \left( G^{+}_{-\half}
\otimes\bbbone\right)\;(\phi\otimes\phi_{0}) \right\rangle}
\nonumber \\
& = &
{\displaystyle
\left\langle \phi', \left(G^{-}_{-\half} \; G^{+}_{-\half}
\otimes\bbbone\right)\;(\phi\otimes\phi_{0}) \right\rangle
- \varepsilon_{1}\;
\left\langle \phi', \left( G^{+}_{-\half}\otimes G^{-}_{-\half}
\right)\;(\phi\otimes\phi_{0}) \right\rangle}
\end{eqnarray}
and (\ref{equa1}) we get
\begin{equation}
0  =
\zeta^{-1} \left[ (\kappa + q)\;(2\,(h' + h_{0} - h) + q_{0} + q - q') +
\mu\, (2\,(\kappa - h' + 2 h) + q' ) \right]
\left\langle \phi', \left(\phi \otimes \phi_{0}\right) \right\rangle.
\end{equation}
Thus a necessary condition
for $\phi'$ to be contained in the tensor product of $\phi$ and
$\phi_{0}$ is that  $[..]=0$. As we know in addition --- from
insertion of $\Delta\,(T_{0})$ --- that $q'=q+q_{0}$, we obtain
a quadratic equation for $h'$,  which has the two solutions
\begin{equation}
\left(j,k \right) \otimes \left(\half,\thalf\right) =
\left(j-1,k\right) \oplus \left(j, k+1\right).
\end{equation}
Similarly, we can use the knowledge of the null-vectors of
$(\thalf,\half)$ to obtain
\begin{equation}
\left(j,k\right) \otimes \left(\thalf,\half\right) =
\left(j+1 , k\right) \oplus \left(j, k-1\right).
\end{equation}
As in the previous section we can use the associativity
and symmetry of the tensor product to deduce from
these conditions
restrictions for the
general fusion rules. We obtain
\begin{equation}
(j_{1},k_{1}) \otimes (j_{2}, k_{2}) =
\sum_{j=\max{(j_{2} - k_{1}, j_{1} - k_{2})}+\half}^{j_{1}+j_{2}-\half}
\left[ \left(j,j-j_{1}-j_{2}+k_{1}+k_{2}\right) \right].
\end{equation}
Again these restrictions reproduce the even fusion rules
of \cite{MSS1}, \cite{MSS2} and were first derived
in \cite{K}. As in the
previous section we find additional fusion rules --- the
two different odd fusion rules of \cite{MSS1}, \cite{MSS2} --- if
we require that instead of (\ref{scaln2})
\begin{equation}
\label{scaln2'}
\left\langle \phi',
\left( G^{a}_{-\half} \otimes \bbbone \right)
\left(\phi_{j,k} \otimes \phi_{\half,\thalf} \right)
\right\rangle =
- \varepsilon_{1} \;
\left\langle \phi',
\left( \bbbone\otimes G^{a}_{-\half} \right)
\left(\phi_{j,k} \otimes \phi_{\half,\thalf} \right)
\right\rangle
\end{equation}
does not vanish, where either $a=+$ or $a=-$.

\section{Conclusions}
\renewcommand{\theequation}{6.\arabic{equation}}
\setcounter{equation}{0}

We have shown in this paper that fusion in conformal
field theory can be understood as a certain
ring-like tensor product of representations of
the chiral algebra. We have proved that this
tensor product is associative and symmetric
up to equivalence. We have also
derived explicit formulae for the action of
the chiral algebra, under which the central
extension is preserved.

Having given a precise meaning to fusion,
deriving the fusion rules is now a
purely algebraic problem, namely to decompose
the tensor product into irreducible representations.
We have demonstrated how this can be done
for the case of the $W_{3}$-algebra and the
$N=1$ and $N=2$ superconformal algebras, thereby
recovering the known restrictions for the fusion rules.
Essentially the same analysis
can be carried through for any chiral algebra, e.~g.\
the $N\geq 3$ NS superconformal algebras or
other $W$-algebras.
We have therefore established a unifying framework within which
all chiral algebras can be treated similarly and the
calculation of the fusion rules is straight forward.

\appendix

\section{Comultiplication property}
\renewcommand{\theequation}{A.\arabic{equation}}
\setcounter{equation}{0}

To prove the comultiplication property for the
three algebras we use the same methods as in \cite{MG}, i.~e.\
we use the Cauchy product formula to obtain a power
series expansion for the product of two power series
and compare these coefficients with the coefficients
of the power series of the product. In the following we
have collected the relevant power series expansions.
\begin{equation}
\sum_{m=1-h}^{n} \left( \begin{array}{c} n+h-1 \\ m+h-1
\end{array} \right)
\zeta^{n-m}    = (\zeta + 1)^{n+h-1}
\end{equation}
\begin{eqnarray}
{\displaystyle \sum_{m=1-h}^{n} \left( \begin{array}{c} n+h-1 \\ m+h-1
\end{array} \right)
\zeta^{n-m} m } & = &
{\displaystyle (1-h) (\zeta+1)^{n+h-1} + \left.
\frac{d}{d\varepsilon}\; (\zeta + \varepsilon)^{n+h-1}
\right|_{\varepsilon=1} } \nonumber \\
& = & {\displaystyle (\zeta+1)^{n+h-2} \left( (1-h)\zeta + n \right) }
\end{eqnarray}
\begin{equation}
\begin{array}{rcl}
{\displaystyle \sum_{m=1-h}^{n}  \left( \begin{array}{c} n+h-1 \\ m+h-1
\end{array} \right)
\zeta^{n-m} m^{2} } & = &
{\displaystyle (\zeta + 1)^{n+h-3} \left( (1 - 2\,h + h^{2})\,
\zeta^{2} \right. }  \\
& &
{\displaystyle \left.
+(h-1 + 3 \,n - 2\,h\,n)\,\zeta + n^{2} \right)}
\end{array}
\end{equation}
\begin{equation}
\begin{array}{rcl}
{\displaystyle \sum_{m=1-h}^{n}  \left( \begin{array}{c} n+h-1 \\ m+h-1
\end{array} \right)
\zeta^{n-m} m^{3} } & = &
{\displaystyle (\zeta + 1)^{n+h-4} \left( (1 - 3\,h + 3\, h^{2} -
h^{3})\, \zeta^{3} \right. }  \\
\vspace{0.2cm}

& &
{\displaystyle
+ (-4 + 7\,h - 3\,h^{2} + 7\,n - 9\,h\,n +
3\,h^{2}\,n)\,\zeta ^{2} }  \\
& &
{\displaystyle
\left. (1 - h - 4\,n + 3\,h\,n +  6\,n^{2} -
3\,h\,n^{2}) \, \zeta \; + n^{3} \right) }
\end{array}
\end{equation}
\smallskip

\begin{eqnarray}
{\displaystyle \sum_{m=1-h}^{\infty} \left( \begin{array}{c} n+m-1 \\ n-h
\end{array} \right)
(-1)^{m+h-1} \zeta^{-(n+m)}} & = &
{\displaystyle \frac{(-1)^{n-h+1}}{(n-h)!}\; \frac{d^{n-h}}{d\zeta
^{n-h}}\; \sum_{l=1}^{\infty} (-\zeta)^{-l} }
\nonumber \\
& = & {\displaystyle (\zeta +1) ^{-(n-h+1)}}
\end{eqnarray}
\begin{eqnarray}
{\displaystyle \sum_{m=1-h}^{\infty} \left( \begin{array}{c} n+m-1 \\ n-h
\end{array} \right)
(-1)^{m+h-1} \zeta^{-(n+m)} m} & = &
{\displaystyle \left( -\zeta \;\frac{d}{d\zeta} - n \right)
(\zeta+1)^{-(n-h+1)}}
\nonumber \\
& = & {\displaystyle (\zeta +1) ^{-(n-h+2)} \left( (1-h)\zeta -n \right) }
\end{eqnarray}
and
\begin{eqnarray}
{\displaystyle \sum_{m=n}^{\infty} \left( \begin{array}{c} m-h \\ n-h
\end{array} \right) (-\zeta)^{m-n}} & = &
{\displaystyle \frac{(-1)^{n-h}}{(n-h)!}\; \frac{d^{n-h}}
{d\zeta ^{n-h}}\; \sum_{l=0}^{\infty} (-\zeta)^{l} }
\nonumber \\
& = & {\displaystyle (\zeta +1)^{-(n-h+1)} }
\end{eqnarray}
\begin{eqnarray}
{\displaystyle \sum_{m=n}^{\infty} \left( \begin{array}{c} m-h \\ n-h
\end{array} \right) (-\zeta)^{m-n}} m & = &
{\displaystyle \left( \zeta \frac{d}{d\zeta} + n \right)
(\zeta +1)^{-(n-h+1)} }
\nonumber \\
& = & {\displaystyle (\zeta +1)^{-(n-h+2)} \left( (h-1)\zeta +n \right). }
\end{eqnarray}

\section{$W_{3}$-calculation}
\renewcommand{\theequation}{B.\arabic{equation}}
\setcounter{equation}{0}

We rewrite the equations (\ref{equai}) and
(\ref{equaii}) using (4.9,\,4.10) of \cite{MG},
namely
\begin{equation}
\label{kappa1}
h'\left\langle \phi',\left( \phi\otimes \phi_{0} \right)\right\rangle
= \zeta \left\langle \phi', \left( L_{-1}\otimes
\bbbone \right) \left( \phi\otimes
\phi_{0} \right)\right\rangle
+ (h+h_{0}) \left\langle \phi',\left( \phi\otimes
\phi_{0} \right)\right\rangle
\end{equation}
and
\begin{equation}
\label{kappa2}
\kappa (\kappa -1) \left\langle \phi',\left( \phi\otimes
\phi_{0} \right)\right\rangle =
\zeta^{2} \left\langle \phi',\left(L_{-1}^{2}\otimes
\bbbone\right) \left( \phi\otimes
\phi_{0} \right)\right\rangle,
\end{equation}
where $\kappa=h'-h-h_{0}\;$,
the formulae (\ref{vir1} - \ref{w31})
and the two null-vector equations to obtain
\begin{equation}
\label{rel18}
0=  2\, h_{0} \zeta  \left\langle \phi',
\left(Q_{-2}\otimes\bbbone\right)
\left(\phi\otimes \phi_{0}\right) \right\rangle
+ 2\, h_{0}\;\left\langle \phi',
\left(Q_{-1}\otimes\bbbone\right)
\left(\phi\otimes \phi_{0}\right) \right\rangle
- 3\, q_{0}\;\kappa\;\zeta^{-1}
\left\langle \phi',\left( \phi\otimes
\phi_{0} \right)\right\rangle
\end{equation}
and
\begin{eqnarray}
\label{rel19}
0 & = &
{\displaystyle (5 h_{0} +1) \;h_{0}\;\;
\left\langle \phi',
\left(Q_{-2}\otimes\bbbone\right)
\left(\phi\otimes \phi_{0}\right) \right\rangle} \nonumber \\
& & {\displaystyle + \left( 12\, q_{0} \; \kappa \; (\kappa-1)
+ 6 \,q_{0} \; (h_{0} +1) \;
(\kappa - h)\right) \;\zeta^{-2}
\left\langle \phi', \left( \phi\otimes
\phi_{0}\right) \right\rangle.}
\end{eqnarray}
Furthermore, we can use
\begin{eqnarray}
\label{rel20}
{\displaystyle q'\;
\left\langle \phi', \left(\phi\otimes
\phi_{0}\right) \right\rangle} & = &
{\displaystyle \left\langle \phi',
\Delta\,(Q_{0})\;
\left(\phi\otimes\phi_{0}\right) \right\rangle} \nonumber \\
& = &
{\displaystyle \zeta^{2} \;\left\langle \phi',
\left( Q_{-2}\otimes\bbbone\right)\;
\left(\phi\otimes\phi_{0}\right) \right\rangle
+ 2 \zeta \;\left\langle \phi',
\left( Q_{-1}\otimes\bbbone\right)\;
\left(\phi\otimes\phi_{0}\right) \right\rangle} \nonumber \\
& &
{\displaystyle + (q +  q_{0})\;
\left\langle \phi',
\left(\phi\otimes\phi_{0}\right) \right\rangle}.
\end{eqnarray}
Eliminating from these last three equations the terms containing
$Q_{-1}$ and $Q_{-2}$ we obtain
\begin{eqnarray}
0 & = &
{\displaystyle \left[
q_{0} \left( 12 \,\kappa (\kappa-1)
+ 6 \,(\kappa - h)\, (h_{0} +1)
+ 3 \, \kappa \, (5 h_{0} +1)
+ h_{0} \, (5 h_{0} +1) \right)
\right. } \nonumber \\
\label{bra1}
& & {\displaystyle \left.
- q'\; h_{0}\;(5 h_{0}+1)
+ q\;h_{0}\;(5 h_{0}+1)
\right]
\left\langle \phi',
\left(\phi\otimes\phi_{0}\right) \right\rangle}
\end{eqnarray}
and thus get the condition, that the bracket $[..]$ must vanish.
However, as we have two unknowns, namely $h'$ and $q'$, we have
to find another condition. To this end we observe that
\begin{equation}
\label{rel1}
\begin{array}{rcl}
{\displaystyle q'\;
\left\langle \phi', \left( Q_{-1}\otimes\bbbone \right)
\left( \phi \otimes \phi_{0} \right) \right\rangle}
\vspace{0.2cm}

& = &
{\displaystyle
\left\langle \phi', \Delta\, (Q_{0})\;
\left( Q_{-1}\otimes\bbbone \right)
\left( \phi \otimes \phi_{0} \right) \right\rangle} \\
\vspace{0.2cm}

& = &
{\displaystyle
(q + q_{0})\; \left\langle \phi', \left( Q_{-1}\otimes\bbbone \right)
\left( \phi \otimes \phi_{0} \right) \right\rangle}  \\
\vspace{0.2cm}

& &
{\displaystyle
+ \zeta^{2}\;
\left\langle \phi', \left( Q_{-2}\;Q_{-1}\otimes\bbbone \right)
\left( \phi \otimes \phi_{0} \right) \right\rangle}  \\
\vspace{0.2cm}

& &
{\displaystyle
+ 2\, \zeta \left\langle \phi', \left( Q_{-1}\; Q_{-1}\otimes\bbbone \right)
\left( \phi \otimes \phi_{0} \right) \right\rangle} \\
\vspace{0.2cm}

& &
{\displaystyle
+ \left( \frac{2}{3}\, h\; + \frac{2}{15}
- \frac{1}{240}\, (22 + 5 c)\;\right)
\left\langle \phi', \left( L_{-1}\otimes\bbbone \right)
\left( \phi \otimes \phi_{0} \right) \right\rangle.}
\end{array}
\end{equation}
Furthermore, we have
\begin{eqnarray}
\label{rel2}
0 & = &
{\displaystyle
\left\langle \phi',
\left( 2\, h_{0} \Delta\,(Q_{-1}) - 3\, q_{0} \Delta\, (L_{-1}) \right)
\left( Q_{-1}\otimes\bbbone \right)
\left( \phi \otimes \phi_{0} \right) \right\rangle}
\nonumber \\
& = &
{\displaystyle
2\, h_{0}\, \zeta \;
\left\langle \phi', \left( Q_{-2}\;Q_{-1}\otimes\bbbone \right)
\left( \phi \otimes \phi_{0} \right) \right\rangle
+ 2\, h_{0}\;
\left\langle \phi', \left( Q_{-1}\;Q_{-1}\otimes\bbbone \right)
\left( \phi \otimes \phi_{0} \right) \right\rangle} \nonumber \\
& &
{\displaystyle
- 3\, q_{0}\;
\left\langle \phi', \left( L_{-1}\;Q_{-1}\otimes\bbbone \right)
\left( \phi \otimes \phi_{0} \right) \right\rangle}
\end{eqnarray}
and
\begin{eqnarray}
\label{rel3}
0 & = &
{\displaystyle
\left\langle \phi',
\left( (5 h_{0}+1)\;h_{0}\;\Delta\,(Q_{-2})
- 12 q_{0}\; \Delta\,(L_{-1})^{2} \right. \right. } \nonumber \\
& &
{\displaystyle
\left. \left.
\hspace*{2.5cm} +\, 6\, q_{0}\,(h_{0} +1)\; \Delta\,(L_{-2})
\right) \left( Q_{-1}\otimes\bbbone \right)
\left(\phi\otimes \phi_{0}\right) \right\rangle} \nonumber \\
& = &
{\displaystyle
(5\, h_{0} + 1)\, h_{0}\;
\left\langle \phi', \left( Q_{-2}\;Q_{-1}\otimes\bbbone \right)
\left( \phi \otimes \phi_{0} \right) \right\rangle
+ 12\, q_{0}\;
\left\langle \phi', \left( L_{-1}^{2}\;Q_{-1}\otimes\bbbone \right)
\left( \phi \otimes \phi_{0} \right) \right\rangle}
\nonumber \\
& &
{\displaystyle
+ 6\, q_{0}\, (h_{0} + 1)\, \zeta^{-1}\;
\left\langle \phi', \left( L_{-1}\;Q_{-1}\otimes\bbbone \right)
\left( \phi \otimes \phi_{0} \right) \right\rangle
+ 18\, q_{0}\, q\, (h_{0} +1)\, \zeta^{-3}\;
\left\langle \phi',
\left( \phi \otimes \phi_{0} \right) \right\rangle} \nonumber \\
& &
{\displaystyle
- 6\, q_{0}\,(h_{0} + 1)\, (h + 1)\, \zeta^{-2}\;
\left\langle \phi', \left( Q_{-1}\otimes\bbbone \right)
\left( \phi \otimes \phi_{0} \right) \right\rangle.}
\end{eqnarray}
Eliminating the terms containing products of $Q$-generators from
(\ref{rel1} - \ref{rel3}) we obtain the relation
$$
\left[ (5h_{0} +1)\left(h_{0}\,(q+q_{0}-q') + 3\,q_{0}(\kappa -1)\right)
+ 6 q_{0} \,(h_{0}+1)\,(\kappa - h - 2)
+ 12\, q_{0}\,(\kappa - 1)\, (\kappa - 2) \right]
$$
$$
\hspace*{5.0cm}
\left\langle \phi', \left( Q_{-1}\otimes\bbbone \right)
\left( \phi \otimes \phi_{0} \right) \right\rangle $$
\begin{equation}
= - \left[ (5\,h_{0} + 1)\, h_{0}\,\kappa
\left(\frac{2}{3}h + \frac{2}{15} - \frac{22 + 5c}{240} \right)
+ 18\, q\, q_{0}\,(h_{0} + 1) \right]\; \zeta^{-1}\;
\left\langle \phi',
\left( \phi \otimes \phi_{0} \right) \right\rangle,
\end{equation}
which together with (\ref{rel19}) and (\ref{rel20}) gives us
the condition
$$
72\, q_{0}^{2}\; \kappa^{2} + \kappa
\left[ 9\, q_{0}^{2}\,(9 h_{0} - 3) + 48\, q_{0} h_{0}\,
(q+q_{0}-q')
+ 2 h_{0}^{2}\, (5 h_{0} + 1) \left( \frac{2}{3}h + \frac{2}{15}
- \frac{22 + 5c}{240} \right)\right]
$$
\begin{equation}
\label{bra2}
+ 6\, q_{0}\, h_{0}\, (q+q_{0}-q')\,(9\, h_{0} - 3)
+ 36\, q\, q_{0}\, h_{0}\, (h_{0} + 1) = 0 .
\end{equation}
We can eliminate $q'$ from (\ref{bra1}) and (\ref{bra2}),
thereby obtaining a cubic equation in $h'$. For each of the
three possible solutions of $h'$ there is precisely one
solutions for $q'$.
\bigskip

\noindent {\bf Acknowledgements}

It is a pleasure to thank my PhD supervisor Peter Goddard
for much advice and encouragement. I am grateful to
Matthias D\"{o}rrzapf and to G\'{e}rard Watts for explaining
the $N=2$ superconformal, resp. the $W_{3}$ algebra to me.
I also acknoledge useful discussions with H. Kausch
and A. Kent.

I am grateful to Pembroke College, Cambridge, for a
research studentship and to the Studienstiftung des deutschen
Volkes for financial support.

\end{document}